%% file: note2.tex
\documentclass[12pt]{article}

\usepackage{latexsym}
\usepackage{epsfig}
\usepackage{psfrag}

\input eqmacros
\input totalmacroe

\begin{document}
 \topmargin 0pt
 \oddsidemargin 5mm
 \headheight 0pt
 \topskip 0mm

 \addtolength{\baselineskip}{0.4\baselineskip}

 \pagestyle{empty}

 \vspace{0.5cm}

\hfill RH-03-2002

\vspace{2cm}

\begin{center}

{\Large \bf A Note on Noncommutative Scalar Multisolitons}

\medskip

\vspace{1 truecm}


\vspace{.2 truecm}


 \vspace{0.7 truecm}
Bergfinnur Durhuus$^a$\footnote{email: durhuus@math.ku.dk} and Thordur 
Jonsson$^b$\footnote{e-mail: thjons@raunvis.hi.is}

\vspace{1 truecm}

$^a$Matematisk Institut, Universitetsparken 5

2100 Copenhagen \O, Denmark

 \vspace{.8 truecm}

$^b$University of Iceland, Dunhaga 3,

107 Reykjavik, Iceland

 \vspace{1.5 truecm}


 \end{center}

 \noindent
 {\bf Abstract.} We prove that there do not exist multisoliton solutions of
noncommutative scalar field theory in the Moyal plane which interpolate
smoothly between $n$ overlapping solitons and $n$ solitons with an infinite separation.

 \vfill

 \newpage
 \pagestyle{plain}
   
Solitons in scalar field theories were first studied in
\cite{gsm1}  in the limit of an infinite
noncommutativity parameter $\theta$
where the kinetic term in the action can be neglected.
The existence of rotationally invariant scalar noncommutative solitons at finite 
but sufficiently large $\theta$ was
proved in \cite{paper} and stability of these solitons was established in
\cite{paper2}.  In
\cite{gopakumar,lindstrom2} the moduli space of the infinite $\theta$
solitons is studied and it is argued that solitons attract
at finite $\theta$ so that all multi-soliton solutions at finite
$\theta$ are rotationally invariant around some point.  
In this paper we place this 
conjecture on a firm foundation.  More precisely, we show that there 
does not exist a family of
solitonic solutions interpolating smoothly between two overlapping solitons
and two infinitely separated ones.  The proof is based on using energy
estimates which indeed show that the energy of two infinitely separated
solitons is greater than that of two overlapping ones.
Various aspects of the theory of
solitons in noncommutative scalar field theories are discussed in   
\cite{zhou,gorsky,lindstrom,solovyov,bak2,matsuo,jackson,volovich}.

Solitons in a noncommutative two-dimensional scalar field theory with 
a potential $V$  are finite energy solutions to the variational 
equations of the energy functional
\beq{5}
S ( \vp )={\rm Tr}\left([a, \vp] [\vp,a^*]) +\theta V(\vp 
)\right),
\eeq
where $a^*$ and $a$ are the usual raising and lowering operators of the
simple harmonic oscillator and $\vp$ is a self-adjoint operator on
$L^2(\bbR^2)$.
We assume as in \cite{paper,paper2} that the potential $V$ is at least
twice continuously differentiable
with a second order zero at $x=0$,
$V(x) >0$ if $x\neq 0$,
and that $V$ has
only one local minimum in addition to the one at $x=0$
located at $s>0$.
For technical reasons we assume that the potential is analytic on a
neighbourhood of the interval $[0,s]$ but we do not believe that this
condition is necessary.
We follow the notation of \cite{paper}.

Let $\phi_1$ and and $\phi_2$ denote the stable one and two soliton
solutions which were constructed for sufficiently large $\theta$ 
in \cite{paper,paper2}.  These solitons have the properties
\beq{6}
\phi_1\to s|0\kt\br 0|,~~\phi_2\to s(|0\kt\br 0|+|1\kt\br 1|), 
\eeq
as $\theta\to\infty$, i.e., they converge to the infinite $\theta$ solitons
of \cite{gsm1}.  Let $U_z$ be the unitary operator
on $L^2(\bbR^2)$ which implements
the translation of the origin in the plane to the point $(x,y)$ and $z=x+iy$.  Then
$|z\kt =U_z|0\kt $ is a usual harmonic oscillator coherent state and the
operator $s|z\kt\br z|$ is interpreted as an infinte $\theta$  soliton located at $z$.
Similarly, one can argue \cite{gsm1} that 
\beq{98}
\phi^{(\infty )}_{2,z}=
s(|z_+\kt\br z_+|+| z_-\kt\br z_-|),
\eeq
where
\beq{99}
|z_\pm \kt={|z\kt\pm |0\kt\over \sqrt{2(1\pm e^{-|z|^2/2})}},
\eeq
is a two soliton solution at infinte $\theta$ with the solitons located at
$z$ and the origin.  Clearly
\beq{44}
(\phi^{(\infty )}_{2,z} - s(|z\kt\br z| +s|0\kt\br 0|)\to 0~~~{\rm as}~|z|\to\infty
\eeq
and
\beq{45}
\phi^{(\infty )}_{2,z} \to s(|0\kt\br 0| +|1\kt\br 1|)~~~{\rm as}~|z|\to 0 .
\eeq
It is natural to interpret the operator
\beq{1hh}
\phi_{1,z}=U_z \phi_1 U_z^*
\eeq
as the finite $\theta$ one soliton solution located at $z$.  The question is whether there
is a finite $\theta$ analogue of $\phi^{(\infty )}_{2,z}$.  In
\cite{gopakumar} it is argued that this is not the case.
The argument is based on computing the energy of the infinite $\theta$
solution using the finite $\theta$ energy functional and observing that the
energy depends on the separation between the solitons.
Here we provide a proof.

Let us assume that there exists a family of solitons, $\phi_{2,x}$, $x\geq
0$, depending smoothly on $x$, such that
$\phi_{2,0}=\phi_2$ and 
\beq{91}
\lim_{x\to\infty} (\phi_{2,x}-\phi_{1,x}- \phi_1)= 0,
\eeq
where the convergence above is in the norm $\|\cdot\|_{2,2}$ defined by
\beq{91x}
\|\phi\|_{2,2}^2= \Tr([a,\phi][\phi,a^*]+\phi^2).
\eeq
Then the energy $E(x)=S(\phi_{2,x})$ is a differentiable function of $x$ and since
$\phi_{2,x}$ is a critical point of $S$ we have $E'(x)=0$ so
$E(x)$ is a constant.  Hence,
\beq{76}
S(\phi_2)=E(0)=\lim_{x\to\infty}E(x)=2S(\phi_1).
\eeq
In order to prove the last equality in \rf{76} we write
\beq{a1}
S(\phi )=K(\phi )+\theta \Tr V(\phi), 
\eeq
where 
\beq{a2}
K(\phi )=\Tr([a,\phi][\phi,a^*].
\eeq
We first show that
\beq{a3}
K(\phi_{2,x})\to 2K(\phi_1)
\eeq
as $x\to\infty$.  By the triangle inequality
\bea
|K(\phi_{2,x})^{1/2}-K(\phi_{1,x}+\phi_1)^{1/2}| & \leq &
K(\phi_{2,x}-\phi_{1,x}-\phi_1)^{1/2}\nonumber\\
  & \leq & \|\phi_{2,x}-\phi_{1,x}-\phi_1\|_{2,2}.
\eea
In order to establish \rf{a3} it therefore suffices to show that
$K(\phi_{1,x}+\phi_1)\to 2K(\phi_1)$ as $x\to\infty$.  We have
\bea
K(\phi_{1,x}+\phi_1) & = & K(\phi_{1,x})+K(\phi_1) +
2 {\rm Re}\,\Tr [a,\phi_{1,x}][\phi_1,a]\nonumber\\
  & = & 2K(\phi_1)+ 2 {\rm Re}\,\Tr [a,\phi_{1,x}][\phi_1,a].
  \eea
If $f$ is the function on $\bbR^2$ which corresponds to $\phi_1$ under Weyl
quantization, then
\beq{a7}
\Tr [a,\phi_{1,x}][\phi_1,a] = {1\over 4\pi}\int_{\bbR^2}
\pa_{\bar{z}}f(z-x)\pa_z\bar{f}(z)\,dzd\bar{z}
\eeq
and the last integral tends to zero as $x\to\infty$ 
by the Riemann-Lebesgue Lemma (this can
also be seen directly since $f$ is a gaussian).

We next prove the convergence 
\beq{a8}
V(\phi_{2,x})\to 2V(\phi_1)
\eeq
as $x\to\infty$ for $V$ a polynomial.  The proof generalizes without
difficulty to the case where $V$ is analytic on a neighbourhood of $[0,s]$
by means of the analytic functional calculus.
Let $\|\cdot \|_2$ denote the usual Hilbert-Schmidt norm.  For $n\geq 2$ we have
\bea
|\Tr \phi_{2,x}^n-\Tr (\phi_{1,x}+\phi_1)^n| & = &
\Tr \left( \sum_{k=0}^{n-1}\phi_{2,x}^k(\phi_{2,x}-\phi_{1,x}-\phi_1)
(\phi_{1,x}+\phi_1)^{n-1-k}\right) \nonumber\\
  & \leq &\sum_{k=0}^{n-1}\|\phi_{2,x}-\phi_{1,x}-\phi_1\|_2\,
\|\phi_{2,x}\|_2^k\,\|\phi_{1,x}+\phi_1\|_2^{n-1-k}\nonumber\\
  & \leq & c \,\|\phi_{2,x}-\phi_{1,x}-\phi_1\|_2,
\eea
for an $x$-indpendent constant $c$ since 
\beq{a9}
\|\phi_{x,1}+\phi_1\|_2\leq \|\phi_{1,x}\|_2+\|\phi_{1}\|_2
= 2 \|\phi_{1}\|_2.
\eeq
On the other hand,
\beq{b1}
\Tr(\phi_{1,x}+\phi_1)^n\to 2 \Tr\phi_1^n
\eeq
as $x\to\infty$ 
since all cross terms between $\phi_{1,x}$ and $\phi_1$ vanish by the
Riemann-Lebesgue Lemma as before.  We have therefore established the
convergence \rf{a8} for any polynomial $V$ with a second order zero at the
origin and this completes our proof of \rf{76}.
Now we return to our main line of argument.

Differentiating the equation $S(\phi_2)=2S(\phi_1)$ with respect to $\theta$
we obtain
\beq{77}
\Tr V(\phi_2)=2\,\Tr V(\phi_1).
\eeq
We claim that the above equality is violated for all sufficiently large
values of $\theta$ and therefore the family $\phi_{2,x}$ does not exist for
large values of $\theta$.

Let $\lambda_0^{(1)}, \lambda_1^{(1)}, \lambda_2^{(1)}, \ldots$ and
$\lambda_0^{(2)},\lambda_1^{(2)},\lambda_2^{(2)},\ldots$ denote the
eigenvalues of $\phi_1$ and $\phi_2$ in the harmonic oscillator basis.  Then
the eigenvalues form decreasing sequences of positive numbers 
and in \cite{paper2} it is proven that
\bea
\lambda_0^{(1)} & = & s-{2s\over\theta V''(s)}+O(\theta^{-2})\\
\lambda_1^{(1)} & = &  {2s\over\theta V''(0)}+O(\theta^{-2})\\
\lambda_0^{(2)} & = &  s - O(\theta^{-2})\\
\lambda_1^{(2)} & = & s -{4s\over\theta V''(s)}+O(\theta^{-2})\\
\lambda_2^{(2)} & = & {4s\over\theta V''(0)}+O(\theta^{-2})
\eea
and all the other eigenvalues are $O(\theta^{-2})$.   
It follows that
\bea
\Tr V(\phi_1) 
& =  & V(\lambda_0^{(1)})+V(\lambda_1^{(1)})+\sum_{n=2}^\infty
V(\lambda_n^{(1)})\\ 
          \!\!\!    =  \lefteqn{V(s) + \left({2s^2\over  V''(s)}+{2s^2\over  V''(0)}\right)
	     {1\over\theta^2} + \sum_{n=2}^\infty
	     V(\lambda_n^{(1)}) +O(\theta^{-3})}.\label{ooo}
\eea
Similarly,
\bea
\Tr V(\phi_2) &  = &  V(\lambda_0^{(2)})+V(\lambda_1^{(2)})+ V(\lambda_2^{(2)})+
\sum_{n=3}^\infty
V(\lambda_n^{(2)})\\  =
\lefteqn{2V(s) + \left({8s^2\over  V''(s)}+{8s^2\over  V''(0)}\right)
             {1\over\theta^2} +\sum_{n=3}^\infty
	                  V(\lambda_n^{(2)}) +O(\theta^{-3})}.\label{ooo1}
			  \eea
In order to estimate the tails of the sums we use the following argument.
There is a number
$b\in (0,s)$ such that $V(x)\leq xV'(x)$ for $0<x<b$.   Taking $\theta $
sufficiently large so that $\lambda_2^{(1)}<b$ and $\lambda_3^{(2)}<b$ we
see that
\bea
\sum_{n=2}^\infty V(\lambda_n^{(1)}) & \leq & \sum_{n=2}^\infty
\lambda_n^{(1)} V'(\lambda_n^{(1)})\\
&\leq & \lambda_2^{(1)} \sum_{n=2}^\infty V'(\lambda_n^{(1)})\\
&\leq & c\theta^{-2}\sum_{n=2}^\infty V'(\lambda_n^{(1)}),
\eea
where $c$ is a constant.

Similarly,
\beq{yy}
\sum_{n=3}^\infty V(\lambda_n^{(2)})\leq  c\theta^{-2}\sum_{n=3}^\infty
V'(\lambda_n^{(2)}).
\eeq
We also have (see \cite{paper,paper2})
\beq{yy1}
\sum_{n=0}^\infty V'(\lambda_n^{(1)})=\sum_{n=0}^\infty V'(\lambda_n^{(2)})=0
\eeq
so
\beq{yy2}
\sum_{n=2}^\infty
V'(\lambda_n^{(1)})=-V'(\lambda_0^{(1)})-V'(\lambda_1^{(1)})=O(\theta ^{-2})
\eeq
and
\beq{yy4}
\sum_{n=3}^\infty 
V'(\lambda_n^{(2))}=-V'(\lambda_0^{(2)})-V'(\lambda_1^{(2)}) 
-V'(\lambda_2^{(2)})=O(\theta ^{-2}).
\eeq
It follows that the tails of the sums in \rf{ooo} and \rf{ooo1} are $O(\theta^{-4})$ and
Eq.\ \rf{77} can be written as
\beq{yy7}
\left({1\over  V''(s)}+{1\over  V''(0)}\right)
             {4s^2\over\theta^2} +O(\theta^{-3})=
	     \left({1\over  V''(s)}+{1\over  V''(0)}\right)
	                  {8s^2\over\theta^2} +O(\theta^{-3})
\eeq
which is impossible for large values of $\theta$.

By a similar argument one can rule out the existence of a multisoliton
solution $\phi_{n,z_1,\ldots ,z_n}$ at large but finite $\theta$ with the
properties
\beq{tt}
\phi_{n,z_1,\ldots ,z_n}\sim s\sum_{i=0}^{n-1}|i\kt\br i|, ~~~z_i\to 0
\eeq
and 
\beq{tt1}
\phi_{n,z_1,\ldots ,z_n}\sim \sum_{i=1}^{n} \phi_{1,z_i},~~~|z_i-z_j|\gg
1,~ i\neq j.
\eeq

The fact that general multisolitons do not exist as solutions to the
static equations of motion at finte $\theta$ will most likely make it more
difficult to establish rigorously the
results about moduli space approximation \cite{lindstrom2}.
To actually control such approximations seems
to require an analysis of the time dependent equations of motion.
 
\medskip

\noindent
{\bf Acknowledgements. } 
The work of B.~D.\ is supported in
part by MaPhySto funded by the Danish National Research Foundation.  
This research was partly supported by TMR grant no. HPRN-CT-1999-00161.
T.~J. is indebted to the Niels Bohr Institute 
for hospitality and would like to thank J. Ambj\o rn, P. Austing and L.
Thorlacius for discussions.

\end{document}

%% file: eqmacros.tex
\def\void{}
\def\labelmark{}

\newenvironment{formula}[1]{\def\labelname{#1}
\ifx\void\labelname\def\junk{\begin{displaymath}}
\else\def\junk{\begin{equation}\label{\labelname}}\fi\junk}%
{\ifx\void\labelname\def\junk{\end{displaymath}}
\else\def\junk{\end{equation}}\fi\junk\labelmark\def\labelname{}}

{\ifx\void\labelname\def\junk{\end{array}\end{displaymath}}
\else\def\junk{\end{array}\right.\end{equation}}
\fi\junk\labelmark\def\labelname{}\def\junk{}
}

\newcommand{\beq}{\begin{formula}}
\newcommand{\eeq}{\end{formula}}
\newcommand{\beqv}{\begin{formula}{}}

%% file: totalmacroe.tex
\newcommand{\rf}[1]{(\ref{#1})}

\newcommand{\bea}{\begin{eqnarray}}
\newcommand{\eea}{\end{eqnarray}}
\newcommand{\beas}{\begin{eqnarray*}}
\newcommand{\eeas}{\end{eqnarray*}}
\newcommand{\beqs}{\begin{displaymath}}
\newcommand{\eeqs}{\end{displaymath}}

\newcommand{\br}{\langle}
\newcommand{\kt}{\rangle}

\newcommand{\vp}{\varphi}

\newcommand{\Tr}{{\rm Tr}\;}
\newcommand{\ben}{\begin{equation}}
\newcommand{\een}{\end{equation}}

\newcommand{\bdm}{\begin{displaymath}}
\newcommand{\edm}{\end{displaymath}}

\newcommand{\pa}{\partial}

\newcommand{\bbR}{{\bf R}}